\documentclass[%
 reprint,
 amsmath,amssymb,
 pra,
floatfix,
]{revtex4-1}

\usepackage{float}
\usepackage{graphicx}
\usepackage{dcolumn}
\usepackage{bm}

\begin{document}
\preprint{APS/123-QED}

\title{Space charge aberrations \\ in single shot time-resolved transmission electron microscopy}

\author{P. Denham}
 \email{pdenham@physics.ucla.edu}
\author{P. Musumeci}%
\affiliation{%
 Department of Physics and Astronomy,
 University of California at Los Angeles, Los Angeles, CA, 90066
}%

\date{\today}

\begin{abstract}
In this paper, we discuss effects of space charge fields on imaging performance in a single shot time-resolved electron microscope (TEM). Using a Green's function perturbation method, we derive analytical estimates for the effects of space charge nonlinearity on the image formation process and the associated aberration coefficients. The dependence of these coefficients on the initial beam phase space distribution is elucidated. The results are validated by particle tracking simulations and provide fundamental scaling laws for the trade-off between temporal and spatial resolution in single-shot time-resolved TEM.
\end{abstract}

\maketitle


\section{\label{Introduction} Introduction}

Transmission electron microscopy (TEM) has proven to be an
extremely powerful and versatile tool in all research areas which benefit from imaging at atomic scale spatial resolution \cite{reimer,spence,williams}. Although a tremendous amount of information can be obtained looking at static snapshots of samples with nm and sub-nm resolution, there is a clear potential for breakthrough advances if the reach of the technique could be upgraded to include the study of how sample structure, composition, and properties change in response to applied stimulus, in other words with the addition of time as a fourth dimension to electron microscopy \cite{zewail}.

On the other hand, in stark contrast with the exceptional progress in  spatial resolution  (recently breaking the sub-angstrom barrier with the introduction of aberration correction \cite{haider:abercorr,krivanek:subangstrom}), the temporal resolution of TEMs is limited due to the intrinsic need of relatively long exposure times to beat the fundamental shot-noise limit of the electron detectors \cite{rose:criterion}. Given average electron currents in TEM columns (typically much below 1 $\mu$A), in order to deliver a illumination dose sufficient to achieve high quality imaging, time intervals on the order of millisecond or longer are required.

There have been multiple attempts to address this shortcoming in electron imaging. One solution is to maintain very low currents in the electron column, but synchronize the time of arrival of the electrons at the object plane with the occurrence of the effect being investigated and repeat the specimen illumination millions of time in the same exact manner\cite{zewail}. This stroboscopic approach has allowed seminal results in the imaging of electric and magnetic field dynamics (PINEM and magnetic vortex)\cite{barwick:pinem,Carbone:lorentzTEM,vanacore:plasmonicvortex}.

When the sample dynamics can not be reproduced in the same way continuously (irreversible processes), one has to resort to single shot illumination, that is send all the electrons in one bunch whose temporal duration sets the exposure time of the microscope\cite{bostanjoglo,LLNL:DTEM}. Incidentally single shot illumination might also be relevant in setups where there are concerns that the electron dose itself would prevent multiple shot accumulation on the detector \cite{spence:outrunningdamage}. Compared to standard TEM operating modes, the peak current in such single-shot illumination regime is necessarily many orders of magnitude higher. For example, 10$^8$ electrons packed in a 10 ps bunch length correspond to peak currents greater than 1 Amp. At these currents, space charge effects quickly degrade the quality of the imaging. Recent work has shown that increasing the energy of the electron beam to the MeV range allows to take advantage of relativistic-induced suppression of space charge forces to restore the microscope spatial resolution \cite{rkli:tem,xiang:tem}.

Two main kinds of space charge interactions, smooth mean field effects and binary collisions, have been identified to play a major role in the evolution of the electron beam distribution in a TEM. Past studies \cite{jansen:coulomb,reed:stochastic, rkli:tem} focused on numerical investigations to elucidate the effects on the image quality of stochastic binary collisions for prototypical TEM column. On the other hand, computer simulations naturally suffer from a loss of generality as only the behavior in one particular optical setup is quantitatively captured. Analytical formulas and scaling laws would allow to quickly estimate spatio-temporal resolution limits for microscope setups spanning many orders of magnitude in  electron energies and peak currents.

In this paper we build an analytical framework to calculate from first principles the smooth mean field space charge induced aberrations extending previous literature (for a good review see \cite{kruit1997space}) to the relativistic regime and validate our formulas against particle tracking simulations \cite{pulsar} which include both smooth space charge and stochastic binary effects separately. In particular, we adopt a perturbation method to study the effects of the smooth space charge field associated with the beam charge density distribution. At first order, the space charge field simply adds a distributed linear defocusing lens over the entire column. At the next higher order, the space charge fields give rise to classical aberration coefficients similar to the ones associated with magnetic lenses, as already recognized in early work by Hawkes and others \cite{hawkes, orloff}. Some of these non linearities simply depend on the initial position on the sample plane and cause a distortion in the final image which could be calibrated in post processing. Other terms affect the trajectories regardless of the initial offset contributing to a blur in the final image plane which amounts to a loss of spatial resolution. One of the main results of this paper is to be able to quantitatively assess these effects and identify the tradeoffs between spatial and temporal resolution for various beam energies. The other important outcome of this study is the identification of possible compensation schemes properly designing the initial beam distribution in the transverse phase space illuminating the sample. In principle, these aberrations can also be corrected by using multipole electron optics in the transport, but the complexity of such systems would quickly become hard to manage as the compensation critically depends on the beam charge distribution.

It is important at this point to note that stochastic Coulomb interaction effects scale like the square root of the charge density in the beam \cite{jansen:coulomb}, while smooth space charge effects increase linearly with the beam current. This suggests that for sufficiently large beam currents (typical for single shot TEM applications), there is always a regime in which the latter (i.e. the subject of this paper) will be dominating the beam dynamics.

The paper is organized as follows: we first review the Green's function approach we will use to evaluate analytically the aberrations and benchmark them with particle tracking simulations in few simple cases where no space charge is present (i.e. for spherical and chromatic aberrations of magnetic round lenses). In this framework, we then move on to the analysis of space charge induced non linearities in the transport and derive handy analytical expressions for the space-charge aberration coefficients. The formulas obtained do depend on the particular shape of the initial beam phase space, but for simple cases (uniform, gaussian) they yield practical estimates for the aberrations enabling quantitative assessment of the trade offs between spatial and temporal resolution for various electron column designs. It is envisioned that the results presented in this paper can guide the development and performance expectations of single shot time-resolved TEM and their scientific application range.

\section{\label{Analytical results} Lens Aberrations}
\begin{figure}[ht]
  \includegraphics[scale=0.45]{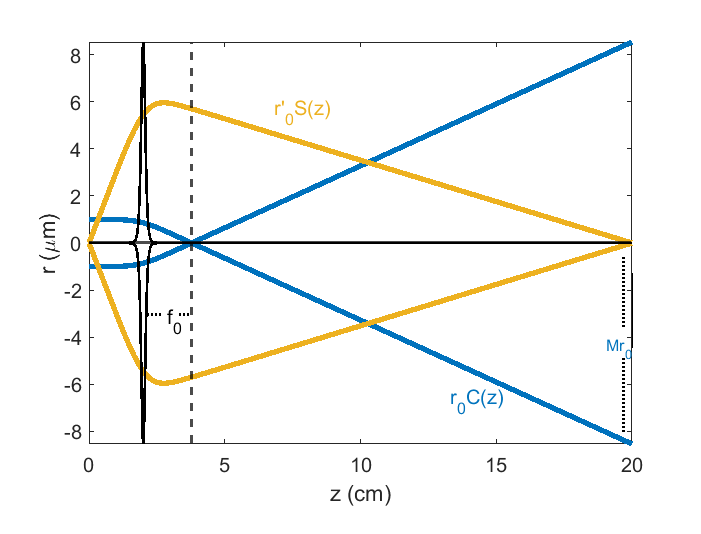}
  \caption{Schematic example of the two principal trajectories in a thin lens imaging stage.}
  \label{Cartoon}
\end{figure}

The unperturbed radial equation for an electron in a cylindrically symmetric (round) lens field with focusing strength
$\kappa(z)$ is given by:
\begin{equation}
r''+\kappa(z)r=0
\label{Hill}
\end{equation}
where $\kappa(z)=\left(\frac{B_0(z)}{2[B\rho]}\right)^2$, $B_0(z)$ is the axial lens magnetic field profile and $B\rho = m_0 c \beta \gamma / e$ the magnetic rigidity of the beam where $m_0$ and $e$ are the electron rest mass and charge and $\gamma$, $\beta$ the usual relativistic factors.
It is well known that the general solution to this equation can be expressed using a symplectic map as:
\begin{equation}
\begin{pmatrix}
r\\
r'
\end{pmatrix}
=
\begin{pmatrix}
C(z) &S(z) \\
C'(z) & S'(z)
\end{pmatrix}
\begin{pmatrix}
r_0\\
r_0'
\end{pmatrix}
\end{equation}
where $C(z)$ and $S(z)$ are two linearly independent solutions to Eq. \ref{Hill}, satisfying $C(0)=1$, $C'(0)=0$, $S(0)=0$, and $S'(0)=1$. Furthermore, since the mapping is symplectic, $C(z)S'(z)-C'(z)S(z)=1$. These two solutions are known as cosine-like and sine-like trajectories. Together, they form a basis in the unperturbed trajectory space (i.e. each trajectory can be written as a linear combination of these functions). The corresponding rays are schematically shown in Fig. \ref{Cartoon}.
When $\kappa(z)$ is designed to image over a distance $L$ , $C(L)=M$ and $S(L)=0$, where $M$ is the magnification. These two conditions ensure the transport map is imaging because the final position is simply a scalar multiple of the initial position independently of the initial angle.

\subsection{Perturbation of trajectories in a model lens}
Throughout the discussion, we will allow the space charge forces whenever present to modify the linear transport, but notably assume that the non linearities associated with deviation from design energy, large angles, space charge, etc. induce only small image plane deviations and the transport can be well approximated by the first order linear optics\cite{Wiedemann:1083415}. In order to evaluate the aberrations, we utilize a Green's function approach to solve the driven Hill equation for the electron transverse motion in a TEM column.

In presence of perturbing forces, we then write the particle trajectory as $r=r_c(z)+\delta r(z)$ where $\delta r(z)$
satisfies
\begin{equation}
\delta r ''+\kappa(z)\delta r = f(z)
\end{equation}
and $r_c(z)$ is the characteristic solution of the Hill equation in the absence of the perturbation $f(z)$.

If we express the driving force as a superposition of impulses, the solution can be written as a convolution integral:
\begin{equation}
\delta r(z) = \int^z G(z,s)f(s)ds
\end{equation}
where the integration interval is from the object to any position along the column up to the image plane and
$G(z,s)$ is the Green's function of the problem which satisfies:
\begin{equation}
\frac{\partial^2G(z,s)}{\partial z^2}+\kappa(z)G(z,s)=\delta(z-s)
\label{Greenfunction}
\end{equation}

Considering separately the cases when $z\neq s$ we can write:
\begin{equation}
G(z,s)=\left\{\begin{matrix}
A_1(s)C(z)+A_2(s)S(z)~{}~{}~{}z<s\\
B_1(s)C(z)+B_2(s)S(z)~{}~{}~{}z>s
\end{matrix}\right.
\end{equation}
as a linear combination of the cosine-like and sine-like basis functions.

Applying the proper boundary conditions at $z=s$ (i.e. continuity of $G$ and discontinuity for the derivative as required by integrating once around the $\delta$-function in Eq. \ref{Greenfunction}) we get:
\begin{equation}
G(z,s) =
  \begin{cases}
    C(s)S(z)-S(s)C(z) & \text{$z>s$} \\
                    0 & \text{$z<s$} \\
  \end{cases}
\end{equation}
This allows us to solve for the excursion from the reference orbits
\begin{equation}
\delta r(z)= \int_0^z\left[C(s)S(z)-S(s)C(z)\right]f(s)ds
\end{equation}

At the imaging plane, $S(L)=0$ and $C(L)=-M$, where $M$ is the magnification of the optical system, so the image plane deviations can be written as:
\begin{equation}
\delta r(L)= M\int_0^LS(s)f(s)ds
\end{equation}


\subsection{Chromatic Aberration}

\begin{figure}[ht]
\centering
  \includegraphics[scale=0.5]{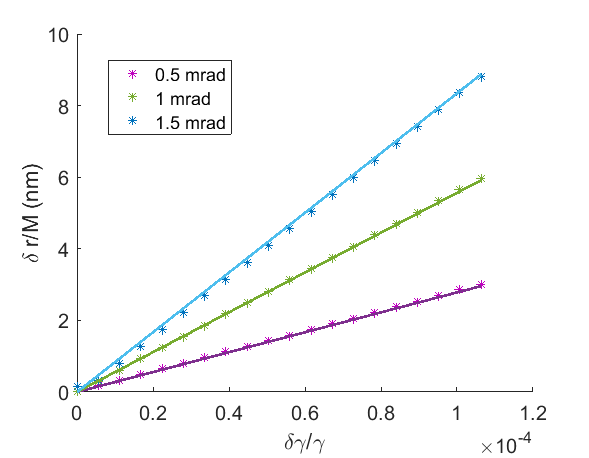}
  \caption{ GPT output of image plane deviations normalized by magnfication and deflection angle and plotted with respect to rms energy spread. }
  \label{chromatic:sims}
\end{figure}

The first example of image plane deviation that we analyze is the one resulting from the chromatic aberration. A particle with an energy slightly higher (lower) than the design value for the optical column will experience a focusing kick slightly weaker (stronger) than the reference particle. For a small relative momentum deviation $\delta p /p$, the corresponding focusing strength acting on this particle can be approximated by:
\begin{equation}
\left(\frac{B_0(z)}{2[B\rho]}\right)^2\left(\frac{1}{1+\frac{\delta p}{p}}\right)^2\approx \left(\frac{B_0(z)}{2[B\rho]}\right)^2\left(1-2\frac{\delta p}{p}\right)
\end{equation}

At first order the equation for the deviation from the reference trajectory can then be written as:
\begin{equation}
\delta r''+\kappa(z) \delta r=2\frac{\delta p}{p}\left(\frac{B_0(z)}{2[B\rho]}\right)^2 r_c(z)
\end{equation}
which can be solved using the Green's function method described earlier. For sine-like reference trajectories $r_c(z)=r_0'S(z)$, the image plane deviation is:
\begin{equation}
\delta r(L)=Mr_0' \frac{\delta p}{p} \int_0^L \frac{S(s)^2}{2} \left(\frac{B_0(s)}{[B\rho]}\right)^2 ds=Mr_0' \frac{\delta p}{p} C_c
\label{chromatic_aberration}
\end{equation}
which can be used to evaluate the chromatic aberration coefficient $C_c$. In Table \ref{parameters} we list the nominal optical and beam parameters for the simulation results presented throughout this paper when not indicated differently. In this analysis, the model equation for the on-axis field of the solenoid lens is:
\begin{equation}
B_0(z)=\frac{\mu_0}{2} \frac{\text{NI}}{d} \left(\frac{z+\frac{d}{2}}{\sqrt{\left(z+\frac{d}{2}\right)^2+R^2}}-\frac{z-\frac{d}{2}
  }{\sqrt{\left(z-\frac{d}{2}\right)^2+R^2}}\right)
  \label{Eq:Bz}
\end{equation}
with physical dimensions of $d$ = 0.015 m, $R$ = 0.008 m, and N=1720. The resulting lens has effective length 1.4 cm, and at around 20 amps, has a focal length of 1.5 cm for a 4.3 MeV beam. The lens images over a distance of 0.2m with a magnification of x8.5.

In Fig. \ref{chromatic:sims} the chromatic aberration coefficient $C_c = 3$ cm obtained from Eq. \ref{chromatic_aberration} is found in excellent agreement with the imaging plane deviation obtained from General Particle Tracer (GPT) \cite{pulsar} simulations performed with space charge effects turned off for various beam divergence angles as a function of the input energy spread. Note that here and in all the plots that follow in the paper the image plane deviations are divided by the magnification factor to relate them to object plane distances. 

\begin{table}[ht]
\caption{\label{parameters} Nominal parameters for single solenoid lens stage GPT simulations.}
\begin{ruledtabular}
\begin{tabular}{lcdr}
\textrm{Parameter}&
\textrm{Value}\\
\colrule
Full Width Pulse Length & 10 ps\\
E-beam kinetic energy & 4.3 MeV\\
E-beam charge & 250 fC\\
Peak Dose at sample & 0.5e/nm$^2$\\
Spotsize/Edge Radius & 1 $\mu$m\\
Beam Divergence &  3 mrad\\
RMS Energy Spread & $<$ 10$^{-5}$\\
Lens Focal Length & 1.5 cm \\
Object to Image plane distance &  20 cm\\
Magnification & 8.5 \\
\end{tabular}
\end{ruledtabular}
\end{table}
At vanishing energy spreads, a small difference between the analytical prediction and the simulation results can be noticed and is due to the fact that the imaging is not perfect even in the absence of chromatic aberrations due to the finite beam divergence and the spherical aberration term which we move on to analyze in the next section.
\subsection{Third Order Lens Aberrations}

Whenever the energy spread of the beam can be kept sufficiently low to minimize the chromatic effects, the main contributions to the trajectory deviation from the ideal imaging condition will be associated with the radial dependence of the focusing field in magnetic round lenses (spherical aberrations). Non linear effects arise due to the longitudinal velocity variation through the lens and higher order terms in the magnetic field components. Following the description in Reiser \cite{reiser} (and again assuming that these terms can be treated as perturbation), we can write for the driven Hill equation:
\begin{widetext}
\begin{align*}
\delta r''+\kappa(z)\delta r&=-\kappa(z)r_c'(z)^2r_c(z)+\kappa(z)\left(\frac{B_0'}{B_0}\right)r_c'(z)r_c(z)^2-\left(\kappa(z)^2-\frac{1}{2}\kappa(z)\left(\frac{B_0''}{B_0}\right)\right)r_c(z)^3 \\
&=f(r_c(z),r_c'(z),B_0(z))
\end{align*}

For a general trajectory having initial position and angle offset $r_c(z)=r_0C(z)+r_0 'S(z)$ the deviation at the image plane due to the radial dependence of the lens field can be written as:
\begin{equation}
\delta r(L)= M\left(r_0^3C_p+3r_0^2r'_0C_q+3r_0r_0^{\prime2}C_r+r_0^{\prime3}C_s\right)
\label{Eq:all_aberrations}
\end{equation}
where the coefficients with subscripts p, q, r and s are related to the classical distortion, coma, image curvature and spherical aberration terms respectively \cite{hawkes}.

The spherical aberration coefficient $C_s$ can be extracted by setting $r_0=0$. Then convolution with the Green function at the imaging plane yields:
\begin{equation}
\delta r(L)=Mr_0^{\prime3}\int_0^LS^4\left(\kappa \left(\frac{B_0'}{B_0}\right)\left(\frac{S'}{S}\right)+\frac{1}{2}\kappa\left(\frac{B_0^{\prime\prime}}{B_0}\right)-\kappa^2-\kappa\left(\frac{S'}{S}\right)^2\right)ds=Mr_0^{\prime3}C_s
\label{Eq:spherical_aberration}
\end{equation}
\end{widetext}

\begin{figure*}[ht]
  \includegraphics[scale=0.5]{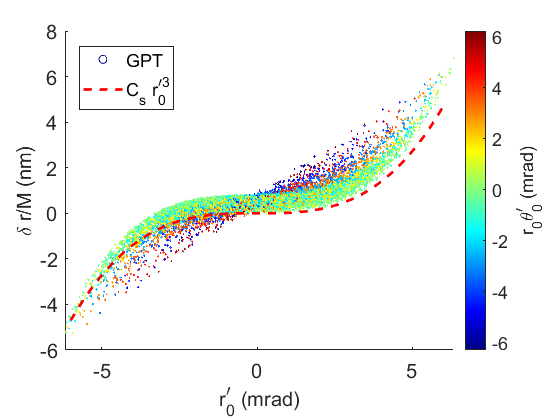}(a)
  \includegraphics[scale=0.5]{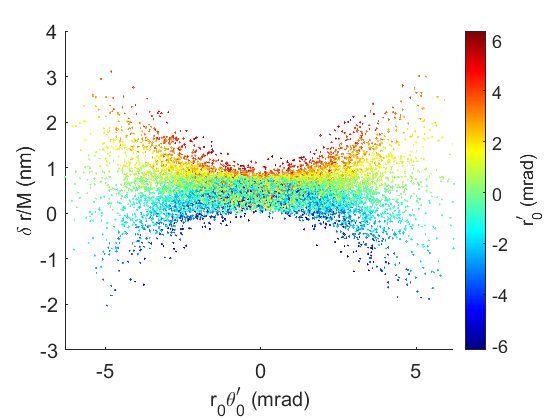}(b)
  \includegraphics[scale=0.25]{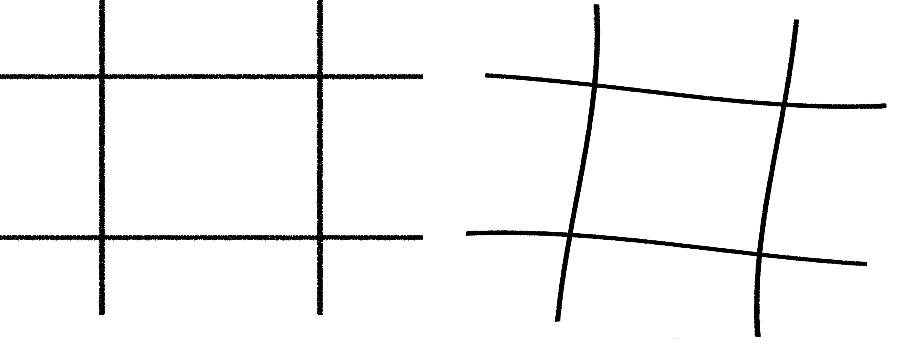}(c)
  \caption{(a) Image plane deviations from the output of a GPT simulation with no energy spread and no space charge effects. The dashed line shows the analytical cubic function from Eq. \ref{Eq:spherical_aberration}. Particles are color-coded by initial angular velocity to show the effect of so called handkerchief aberration. (b) Image plane output plotted against $r_0\theta^{\prime}$ after removing spherical aberration cubic dependence. (c) GPT simulation displaying handkerchief distortion of 10mm hash symbol with 125um bar width.}
  \label{Fig:spherical_aberration}
\end{figure*}

In Fig. \ref{Fig:spherical_aberration} we show a comparison for the image plane deviations obtained using the analytical results from Eq. \ref{Eq:spherical_aberration} and the numerical GPT simulation for a monochromatic beam with no space charge (and other parameters as listed in Table \ref{parameters}). The beam distribution at the object plane is assumed uniform within a 1~$\mu$m hard-edge radius and a 3~mrad rms gaussian angular spread. The excursions from the ideal reference trajectory plotted as a function of the initial ray deflection show an excellent agreement between the calculated $C_s = 3$ cm cubic dependence and the particle tracking output. As an example, for incidence angles of 3~mrad, the particular lens employed in our simulation (i.e. Eq. \ref{Eq:Bz}) contributes to a blurring in the image plane of 0.81~nm. Similarly to the chromatic aberration coefficient $C_c$, it is also typical for $C_s$ to be comparable with the lens focal length. Appropriately crafting the longitudinal on axis magnetic field profile can reduce the spherical aberration.

A careful reader would notice that while the cubic dashed line well reproduces the general behavior of the perturbed trajectories, there is an additional broadening of the image deviations obtained from the particle tracking simulation. This is due to another type of lens aberration and deserves a separate discussion. In our simple treatment of the electron dynamics in the cylindrically symmetric column, we have neglected the azimuthal motion, but due to the solenoid Larmor rotation combined with the radial aberrations in Eq. \ref{Eq:all_aberrations} can result in a deviation in the image plane proportional to the square of the particle initial angular velocity $r_0\theta_0'$ (indicated by the color coding in the left figure and more explicitly in Fig.\ref{Fig:spherical_aberration}(b) which shows the image plane deviation with the cubic spherical aberration subtracted). We also show the results of a GPT simulation aimed at elucidating this effect (characteristically called handkerchief aberration) in Fig.\ref{Fig:spherical_aberration}(c). Here a very large field of view is used so that the handkerchief distortions are clearly visible in the image plane. In general this effect can be neglected if one is interested in the imaging performances limit at very small transverse offsets from the optical axis.

\section{Space Charge Aberration}
In this section, we apply the theoretical framework we developed above to estimate the third order space charge induced deviation from the the unperturbed trajectories.

For a very long electron bunch of length $L_b$ (i.e. having a cigar aspect ratio in the beam rest-frame), where the input charge density satisfies $\frac{1}{\rho}\frac{\partial{\rho}}{\partial z}\ll \frac{1}{L_b}$, the space charge field is predominantly 2D and the longitudinal components can be effectively neglected. In this case, for a known charge density distribution and a cylindrically symmetric optical system, the transverse electric field can be derived from a simplified form of Gauss' law:\begin{equation}
    E_r(r;z)=\frac{1}{r}\int_0^r\frac{\rho_{sc}(\xi;z)}{\epsilon_0} \xi d\xi
    \label{Eq:spacechargefield}
\end{equation}
where $\xi$ is radial integration variable, and the $z = c\beta t$ dependence in Eq. \ref{Eq:spacechargefield} parametrizes the evolution of the transverse density. Specifically, it represents the average position of the charge distribution in the TEM column. Our strategy for calculating third order space charge aberrations proceeds as follows: we start from calculating the transverse charge density evolution using the method of characteristics. Once the evolved transverse density is known, the first and third order space charge fields are computed using Gauss’ law. Finally, the third order field is weighted by the Green’s function and integrated over the column to obtain deflections of the linear trajectories in the image plane and the corresponding aberration coefficients.




Assuming the non linear space charge forces only account for a small perturbation on the motion, the particles in the beam will all evolve along characteristic orbits which can be written in the form of a linear transport map $w=Rw_0$, where $w=(x,x',y,y')^T$ and the 4x4 symplectic matrix $R$ describes the linear uncoupled dynamics in the Larmor frame:
\begin{equation}
R=\begin{bmatrix}
C & S & 0 & 0\\
C' & S' & 0 & 0\\
0 & 0 & C & S\\
 0& 0 & C' & S'
\end{bmatrix}
\end{equation}
The dynamics are Hamiltonian, so the initial distribution is stationary in phase space. Thus, at any given position along the optical column, the distribution function can be written as $f(R^{-1}w) = f(w_0)$ where $f(w_0)$ is the initial distribution. The space charge density at any position in the optical column can then be computed performing the integral over the momentum space:
\begin{multline}
\rho_{sc}(x,y;z) =\iint  dx' dy' f(R^{-1}(x,x',y,y')) \\
=\iint  \frac{dudv}{C(z)^2}f\left(\frac{x-S(z) u}{C(z)},u,\frac{y-S(z) v}{C(z)},v\right)
\end{multline}
where the substitutions $u=C(z) x'-C'(z) x$ and $v=C(z) y'-C'(z) y$ were used.


In order to calculate the lowest order space-charge induced correction terms in cylindrical symmetry we expand the space charge density in a Maclaurin series:
\begin{equation}
\rho_{sc}(r;z)=\sum_{n=0}^{\infty}\rho^{(2n)}(z)\frac{r^{2n}}{(2n)!}
\end{equation} where the sum only runs through even indices as required by the symmetry and $\rho^{(2n)}$ is the 2$n$ radial derivative of the space charge evaluated on the optical axis.
The explicitly indicated dependence on $z$ is due to the evolution along the beamline of the $C(z)$ and $S(z)$ transport functions.

Substituting the charge density into Eq. \ref{Eq:spacechargefield}, and performing the integral yields:
\begin{equation}
E_r(r;z)=\rho^{(0)}(z)\frac{r}{2\epsilon_0}+\rho^{(2)}(z)\frac{r^3}{8\epsilon_0}+\mathcal{O}(r^5)
\end{equation}
where we truncate the sum after the third order term.

Similarly to what we did in the previous section, the space charge aberration can be calculated from the convolution of the non linear field evaluated along the reference trajectory with the Green's function of the driven Hill equation:
\begin{equation}
\delta r(L)=\frac{eM}{\gamma^3mc^2\beta^2}\int_0^L\rho^{(2)}(z)\frac{r_c(z)^3}{8\epsilon_0}S(z)dz
\end{equation} where the relativistic $\gamma^3$ term in the denominator takes into account the effects due to the relativistic mass increase and to the beam magnetic field forces and the electron longitudinal velocity $\beta$ is used to transform the radial acceleration time-derivatives into spatial derivatives. Note that the effect of the linear space charge defocusing field can be included properly modifying the $C(z)$ and $S(z)$ linear transport trajectories (and adjusting the lens strength to maintain the imaging condition).

Substituting  $r_c(z)=r_0C(z)+r'_0S(z)$  yields the space-charge induced image plane deviation terms similar to Eq. \ref{Eq:all_aberrations}:
\begin{multline}
\frac{\delta r_{sc}(L)}{M}= r_0^3C_e^{(p)}+3r_0^2r'_0C_e^{(q)}+3r_0r_0^{\prime 2}C_e^{(r)}+r_0^{\prime3}C_e^{(s)}
\end{multline}
with the aberration coefficients explicitly given by:
\begin{equation}
C_e^{(p)}=\frac{e}{8\epsilon_0\gamma^3mc^2\beta_0^2}\int_0^L \rho^{(2)}(z)C(z)^3S(z)dz
\end{equation}\begin{equation}
C_e^{(q)}=\frac{e}{8\epsilon_0\gamma^3mc^2\beta_0^2}\int_0^L \rho^{(2)}(z)C(z)^2S(z)^2dz
\end{equation}
\begin{equation}
C_e^{(r)}=\frac{e}{8\epsilon_0\gamma^3mc^2\beta_0^2}\int_0^L \rho^{(2)}(z)C(z)S(z)^3dz
\end{equation}
\begin{equation}
C_e^{(s)}=\frac{e}{8\epsilon_0\gamma^3mc^2\beta_0^2}\int_0^L \rho^{(2)}(z)S(z)^4dz
\end{equation}


\subsection{Space charge aberrations in the uniform illumination case.}
We will first study the behavior of the aberration coefficients in the particular case in which the sample is uniformly illuminated with a beam having a Gaussian spread in angles.
If the beam is focused to a waist at the specimen location, then the initial phase space distribution can be written as:
\begin{equation}
f(x,x',y,y')=
  \begin{cases}
    \frac{Q}{\pi R_0^2 L_{b}}\frac{1}{2\pi \sigma_{\theta}^2}\exp\left(-\frac{(x_0'^2+y_0'^2)}{2\sigma_{\theta}^2}\right) & \text{$r_0^2<R_0^2$} \\
    0 & \text{$r_0^2>R_0^2$} \\
  \end{cases}
\end{equation}
Where $Q, R_0, \sigma_{\theta}$ and $L_{b}$ are the beams total charge, initial edge radius, rms beam divergence in the object plane, and bunch length respectively.

Using the results in the previous section, we can invert the transport map, substitute the initial coordinates in terms of the final coordinates into the distribution function, then integrate over the momentum space to find the charge density evolution along the optical transport.
The charge density and its second derivative evaluated on axis are given by:
\begin{equation}
\rho^{(0)}(z)=\frac{Q}{\pi \sigma_{\theta}^2 S^2 L_{b}}\frac{1-\exp\left({-p^2/2}\right)}{p^2}\end{equation}
\begin{equation}
\rho^{(2)}(z)=-\frac{Q\exp{(-p^2/2) }}{2 \pi\sigma_{\theta}^4S^4 L_{b}}
\end{equation} where $p=\frac{R_0 C(z)}{\sigma_{\theta} S(z)}$. Ultimately, the factor $p$ is a proxy which indicates whether or not the transverse spatial distribution has transitioned to being Gaussian or uniform. For large $p$, the initial spatial distribution dominates and the beam profile is effectively uniform. On the other hand, when $p$ is small it means that the beam distribution is initial angle-dominated and takes a more Gaussian spatial profile, in which case, we expect the space-charge field to exhibit a stronger non-linear character.

\begin{figure*}[ht]
\centering
\hspace*{0cm}
  \includegraphics[scale=0.5]{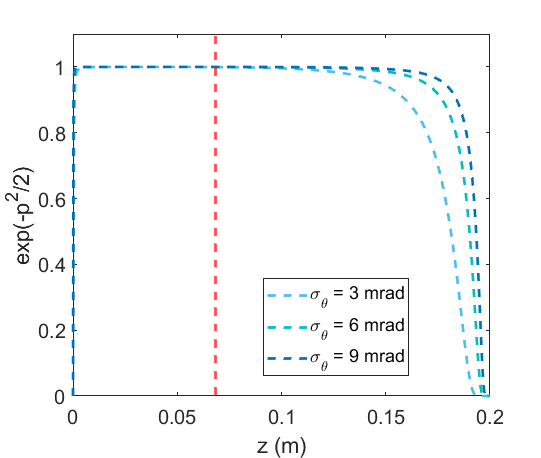}(a)
  \includegraphics[scale=0.5]{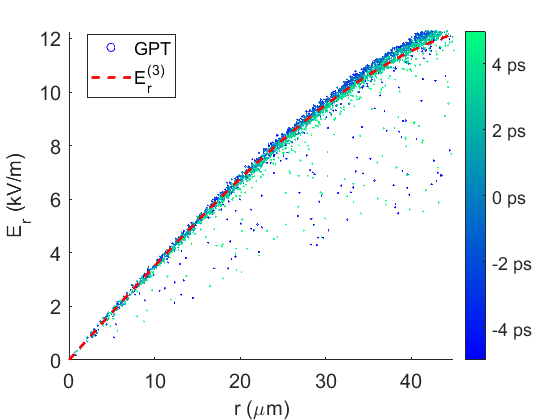}(b)
  \caption{(a) Non linearity coefficient of the charge density distribution plotted from object plane to image plane. At the sampled position along the column indicated by the red vertical dashed line in (a), the radial electric field experienced by each particle as a function of its radial position in the beam is compared with the analytical estimate we used to obtain the space charge aberration coefficients (b). The particles are color coded with respect to their longitudinal position in the beam.}
  \label{p_evolution}
\end{figure*}

We start by noting that in the limit that $p\to 0$ (i.e., when $C\to 0$) the expression $\frac{1-\exp\left({-p^2/2}\right)}{p^2}\to 1/2$, and the zero order density $\rho^{(0)}\to \frac{Q}{2\pi S^2 \sigma_{\theta}^2L_{b}}$, i.e, the on axis density of a Gaussian profile with rms spot size given by $S\sigma_{\theta}$. Alternatively, when $S\to 0$ such as at a object or image plane, $\rho^{(0)}\to \frac{Q}{\pi R_0^2 C^2L_{b}}$, which is the density of a uniformly charged cylinder.

Knowing the second derivative of the space charge density on axis, we can then use the formulas from the previous section to determine the aberration coefficients.
\begin{equation}
C_e^{(p)}=-\frac{K}{8\sigma_{\theta}R_0^3}\int_0^L p^3 \exp{\left(-p^2/2\right)}dz
\end{equation}\begin{equation}
C_e^{(q)}=-\frac{K}{8\sigma_{\theta}^2R_0^2}\int_0^L p^2 \exp{\left(-p^2/2\right)}dz
\end{equation}
\begin{equation}
C_e^{(r)}=-\frac{K}{8\sigma_{\theta}^3R_0}\int_0^L p \exp{\left(-p^2/2\right)} dz
\end{equation}
\begin{equation}
C_e^{(s)}=-\frac{K}{8\sigma_{\theta}^4}\int_0^L\exp{\left(-p^2/2\right)}dz
\label{Eq:spacecharge_aberr}
\end{equation}
where the perveance factor $K = \frac{2I}{I_A\gamma^3\beta^3}$, $I=Qc\beta/L_b$ is the beam current, and $I_A\approx 17 kA$ is the Alfven current. 

For small offsets of the trajectories from the axis, or in other words, if we restrict the field of view to a very small area around the axis, the dominant term in the image plane deviations will be the one associated with Eq. \ref{Eq:spacecharge_aberr}. 

For a setup where $\frac{R_0}{\sigma_\theta f}\ll1$, the beam spends a majority of the time as a Gaussian, and $\int_0^L\exp{\left(-p^2/2\right)}dz\approx L$. For larger initial spot sizes (or smaller focal waist sizes) the beam will have a Gaussian profile only in a region more localized around the back focal plane of the lens.
For the setup corresponding to the electron beam and lens parameters in Table \ref{parameters}, the approximation is justified in Fig. \ref{p_evolution} where $\exp(-p^2/2)$, is plotted for three different beam divergences. In these cases, we have $I$=25~mA, $K=3.6\times10^{-9}$, and $C_e^{(s)}$ = 1~m, much larger than any other aberration contribution so we expect this effect to be dominant at the imaging plane. The sign of the space charge spherical aberration term is the same as the lens spherical aberrations. Moreover, the linear space charge forces modify $C(z)$ and $S(z)$, which in turn increases the value of the lens spherical aberration. However, as will be shown, the third order space charge effects still dramatically overtake the third order lens effects by nearly 2 orders of magnitude for a beam current of 25 mA.

%

\begin{figure*}[ht]
    \centering
    \includegraphics[scale=0.45]{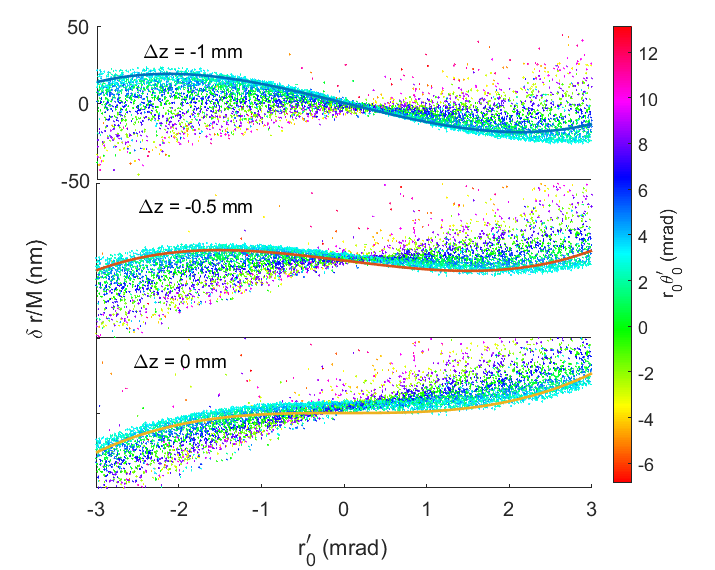}(a)
    \includegraphics[scale=0.425]{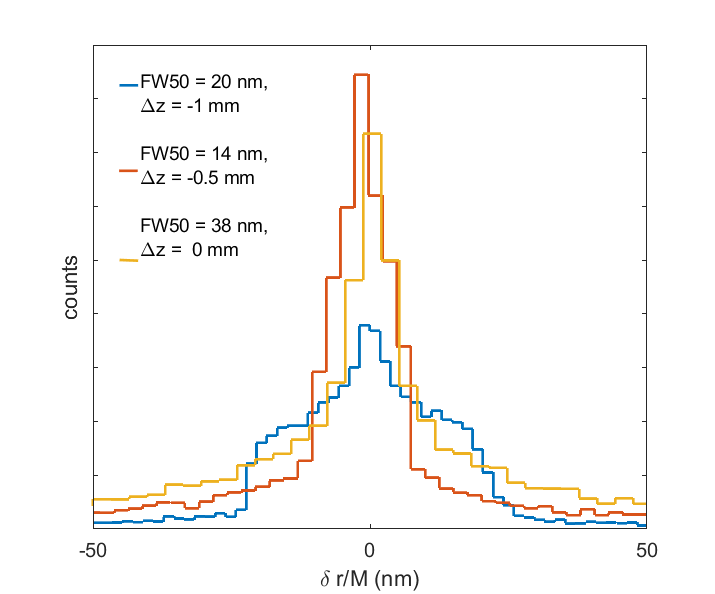}(b)
    \caption{(a) GPT output of image plane deviations and $\frac{\delta r}{M}=C_e^{(s)}r_0'^3+\frac{\Delta z r_0'}{M^2}$, with M$\approx8.5$ and $C_e^{(s)}\approx 1$m, plotted with respect to the initial angle for three different defocus conditions $\Delta z = -1$ mm,  $-0.5$ mm and 0 mm. The latter corresponds to the linear transport image plane. (b) Corresponding histograms for each defocus position.}
    \label{fig:unibeam}
\end{figure*}

\begin{figure}
    \centering
    \includegraphics[scale=0.48]{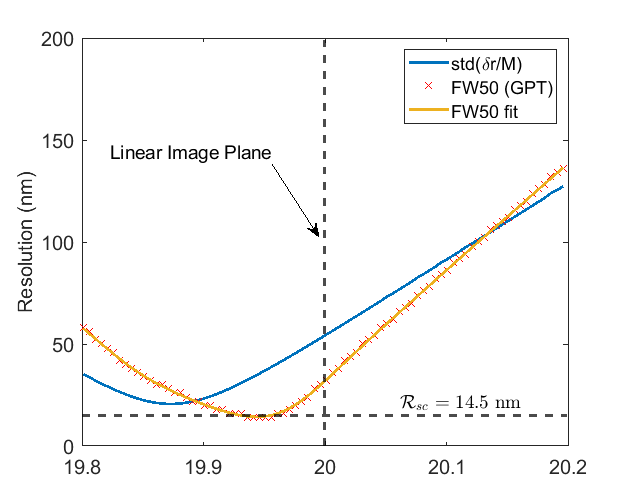}(a)
    \caption{FW50 and standard deviation of image plane excursions plotted with respect to image plane placement. Linear transport is tuned so that the screen at 20 cm exactly satisfies the imaging condition ($\Delta z = 0$).}
    \label{fig:Defocus_scan}
\end{figure}

Once again we use GPT to validate our model and estimates for $\delta r_{sc}/M$. In Fig. \ref{p_evolution}b we show the transverse electric field felt by the particles at an arbitrary location along the beamline (indicated by the dashed line in Fig. \ref{p_evolution}a) compared with the third order polynomial calculated from our analytical model. The agreement is excellent and only particles at the head and tail of the beam (where the assumption of infinite/very long beam breaks down) experience a field deviating significantly from the prediction. The GPT simulation results for the image plane deviations are shown in Fig. \ref{fig:unibeam} in which the overall trend follows the prescribed cubic dependence. For example, using the parameters in Table \ref{parameters}, for particles incident on the sample with angles of 3 mrad, we expect a $30$nm deviation from the perfect imaging condition. Comparing the y-scale with the one in Fig. \ref{Fig:spherical_aberration} indicates how severe can be the effect of space charge induced aberrations on the instrument spatial resolution. The color map highlights the correlation with $r\theta'$ in the object plane discussed above in the context of the handkerchief aberration.

The spatial resolution of the instrument can be estimated by the width of the histogram of the image deviations. Before we move forward, it is important to note that by optimizing the linear defocus of the lens, the width of the projected distribution can be reduced by nearly a factor of 2 with respect to its value at the image plane (see Fig. \ref{fig:Defocus_scan}).


This essentially follows from the Scherzer defocus theorem \cite{scherzer1949}. We can understand what happens more quantitatively by considering the effect on the image plane deviations when moving the output plane back by a small amount $\Delta z$. Considering the spherical aberration term, the output plane deviations are given by:
\begin{equation}
\delta r = \frac{\Delta z}{M}r_0'+MC_e^{(s)}r_0'^3
\end{equation}
In the absence of defocus (i.e. $\Delta z = 0$) and assuming a gaussian angular distribution, the rms spread of the image plane deviations is given by:
\begin{equation}
\sqrt{\langle\left(\delta r/M\right)^2 \rangle} = \sqrt{15}C_e^{(s)}\langle r_0'^2\rangle^{\frac{3}{2}}
\end{equation}
However, it is possible to offset the cubic contribution due to the spherical aberration using a small defocus $\Delta z = -3 M^2 C_e^{(s)}\langle r_0'^2\rangle$ for which we have:
\begin{equation}
\sqrt{\langle \left(\delta r/M\right)^2 \rangle}_{min}=\sqrt{6}C_e^{(s)}\langle r_0'^2\rangle^{\frac{3}{2}}
\end{equation}
thus minimizing the rms spread of the image deviations by a factor of $\sqrt{\frac{6}{15}}=0.6325$.

While rms is the most common quantity to estimate the beam width in accelerator physics, it disproportionately weighs outliers in the distribution. For this reason, in TEM literature it is common to use the FW50 (full width containing 50 $\%$ of the beam distribution). The minimum of this quantity in our example is 14 nm, significantly smaller than the rms width at the image plane as shown in the defocus scan shown in Fig. \ref{fig:Defocus_scan}.

Based on all of these these considerations, in the following section we will use as an approximation for the instrument spatial resolution which is 
\begin{equation}
\mathcal{R}_{sc}(I,\gamma,L,\sigma_\theta) \approx \frac{1}{2}C_e^{(s)} \sigma_\theta^3 = \frac{K L}{16 \sigma_{\theta}}
\end{equation}
which yields 14.5 nm in our example in close agreement with the GPT simulation. Note that if the lens is sufficiently thin, and the space charge modified magnification is close enough to the zero-charge magnification, then $L \approx M f_0$ , where $f_0$ is the lens focal length, suggesting that space charge aberrations would be reduced for smaller focal length optics.

In terms of beam kinetic energy scaling, the perveance is proportional to $\gamma^{-3}$ and for a fixed magnetic field profile, the focal length scales as $\gamma^2$, so the space charge contribution to the overall resolution will scale as $I M/\gamma\sigma_{\theta}$. On the other hand, this assumes that the focal length of the lens can be arbitrarily reduced (i.e. proportional to $\gamma^2$) which might not be fully feasible depending on the magnet technology employed. 

\section{Trade-offs between spatial and temporal resolution}

\begin{figure*}[ht]
\includegraphics[scale=0.5]{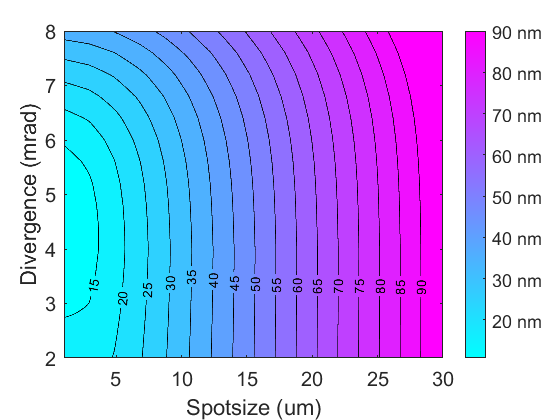}(a)
\includegraphics[scale=0.5]{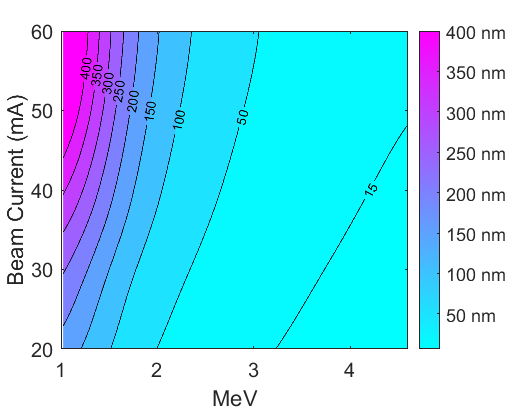}(b)
\includegraphics[scale=0.5]{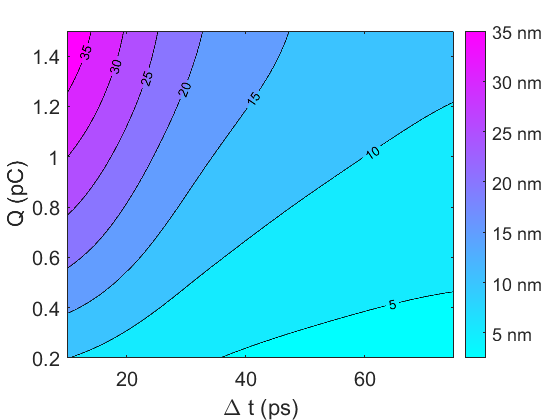}(c)
\includegraphics[scale=0.5]{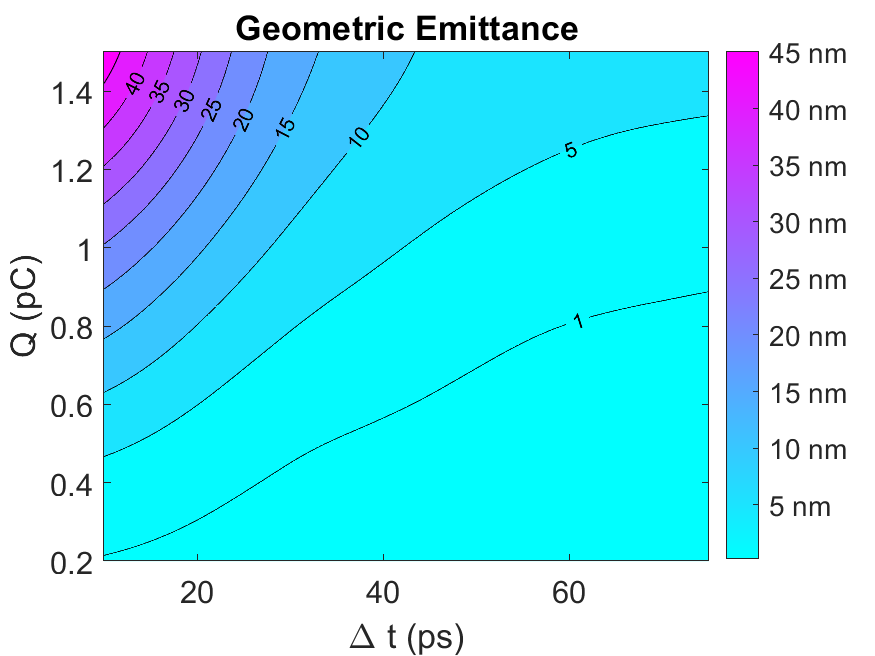}(d)
    \caption{Analytical/numerical matrix propagator calculations of resolution. (a)Scanned dependence of resolution on illumination geometry for 4.3 MeV kinetic energy and 25 mA peak current. (b) Optimal resolution as a function of beam current and energy. The lens, image and object plane position are held constant in this plot. For each energy and beam current, the illumination parameters are chosen to optimize the resolution. 
    (c) Optimized resolution for varying beam charge and bunch length for a 4.3 MeV kinetic energy beam. The illumination geometry is optimized at each point and the corresponding required geometric emittance is shown in (d). }\label{fig:optimum}
\end{figure*}

With this analytical estimate of the space charge aberration it is now possible to estimate the spatial resolution of a time-resolved single-shot transmission electron microscope as a function of the beam energy, beam current, spot size at the sample and maximum opening angle. The cumulative effect of all competing aberrations caused by spherical, chromatic, space charge, and dose resolution limits is much worse than the diffraction limit $R_d=1.22\frac{\lambda}{\theta}$ where $\lambda$ is the electron De Broglie wavelength, so we exclude the contribution due to this term in the estimate.
Notably, when beam energy is larger than 700~keV, the electron De Broglie wavelength is less than 1 pm, so a semicollection angle no larger than a 5 mrad will have a diffraction limit smaller than a Bohr radius.

Assuming independent contributions, the overall resolution can be estimated by the quadrature sum of all the different sources of excursion from the ideal linear imaging condition along with the resolution limit set by the illumination of the specimen as:
\begin{equation}
\mathcal{R}=\sqrt{\left( C_c\frac{\delta \gamma}{\gamma}\sigma_\theta\right)^2+\left(C_s \sigma_\theta^3\right)^2 + \mathcal{R}_{sc}^2+\frac{SNR}{Dose}}
\label{Eq:resolution_estimate}
\end{equation}
where $SNR = 5$ is the desired signal to noise ratio, and $Dose$ is the particle density at the object plane. This latter term simply indicates that if the electron beam charge is too low, there are just not enough electrons in a resolution pixel to statistically resolve low (20 $\%$) contrast features in the image.

The expression for $\mathcal{R}$ in Eq. \ref{Eq:resolution_estimate} can be used as a first approximation to the resolution of a single shot time-resolved TEM useful to understand the tradeoffs between the different parameters, but does not take into account the correlations between the deviations from the imaging condition (they are not all independent). Detailed numerical simulations would still be required to assess the ultimate resolution limit. In this section we consider Eq. \ref{Eq:resolution_estimate} essentially as a multivariate cost function, to be minimized within some reasonable domain of beam and lens parameters to yield the best imaging performances.

For example, the spot size and beam divergence of the beam waist at the sample plane can be optimized using the condenser lens to improve the resolution. Considering the space charge aberration scaling in Eq. \ref{Eq:spacecharge_aberr}, and the inverse dependence on beam divergence makes it clear that increasing beam divergence helps to reduce $\mathcal{R}_{sc}$ and therefore improve resolution up to the point where spherical aberrations begin to dominate. Similarly, increasing the illumination spot size improves the space-charge resolution, but it simultaneously lowers the illumination dose, eventually degrading the imaging performances due to the Rose criterion. An optimal trade off can be found for given peak current and beam energy. 
In practice spot size and divergence at the sample are not independent for a given beam emittance from the electron gun. It is important at this regard to note that the contrast mechanism typically used for imaging is to intercept with an aperture the scattered electrons. Therefore a very large angular divergence at the sample will significantly decrease contrast since as the scattered electron distribution gets mixed with the transmitted electron distribution, the object features become harder to distinguish. A lower beam emittance in this case would provide superior contrast thanks to the lower intrinsic beam divergence.


In Fig. \ref{fig:optimum}(a) we show the estimate for the single shot TEM resolution plotted for the 4 MeV energy and 25 mA peak current as a function of spot size and divergence at the object plane. To generate this plot, the divergence is scanned between 2-8~mrad, and the spot size is varied from 1~$ \mu$m to 30~$\mu$m. The resulting contour map (obtained simply plotting Eq. \ref{Eq:resolution_estimate}) indicates that nearly $10$nm resolution can be achieved if the beam is focused to $1\mu m$ spot size with $\sigma_{\theta}$ = 4~mrad. GPT simulations are in excellent agreement as analyzing the image plane deviations for this optimized illumination geometry yield a FW50 resolution of 12~nm.

In Fig. \ref{fig:optimum}(b), we show the optimum resolution as a function of beam energy and beam current. This is particularly interesting as there are many different electron sources being considered for ultrafast TEM operation \cite{lessner,musumeci:advances, rosenzweig:cryoRF, SRFwigun, yang:microscopy} and an even rough estimate of the space charge aberration can be useful to quickly assess potential performances of a proposed TEM. Higher beam energies are clearly favored here, but it should be pointed out that in this plot we kept the focal length constant to a value of 1.5 cm, and the magnet technology to obtain short focal lengths for higher electron energies becomes increasingly more challenging.

Optimization of the illumination geometry can also be performed for different pulse lengths and charge as long as the beam aspect ratio remains large enough to satisfy the 2D limit approximation. In Fig. \ref{fig:optimum}(c),  the illumination geometry is optimized for various bunch length and charge. The beam kinetic energy in this plot is kept constant at 4.3 MeV. As expected, lower beam charges and longer pulse lengths both improve the resolution. One aspect to keep in mind is that beam divergence and illuminated area at the sample plane are different (and re-optimized) at each point in the plot, so for example a higher beam charge allows to look at a larger field of view albeit with lower spatial resolution. This is not surprising as in order to keep the dose above the Rose criterion, the beam must be focused to a smaller spot size, while the aberrations determine the optimal beam divergence. In principle, sub-5~nm resolution can be achieved using bunch lengths longer than 100 ps and emittances smaller than 5 nm.

Finally, in agreement with what was found through numerical simulation already in \cite{rkli:tem}, it is important to note here that additional magnification stages would have much smaller contributions to the smooth space charge aberrations. Space-charge
aberration coefficients in the second stage in our example are up to 10 times larger than those of the first stage. But due to the $M$ times smaller divergence at the entrance to the second stage (hence $M^3$ times smaller image disk size scaling) and $M$ times larger “object” size, smooth space-charge effects have a negligible impact on image quality in the second and following stages of the column.

\section{Mitigation effects. Reshaping the Distribution.}
\begin{figure*}[ht]
\includegraphics[scale=0.5]{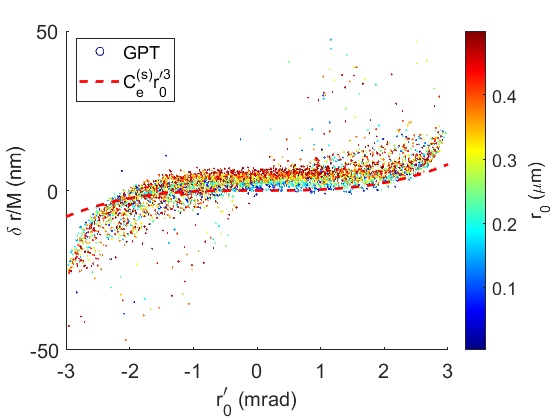}(a)
\includegraphics[scale=0.5]{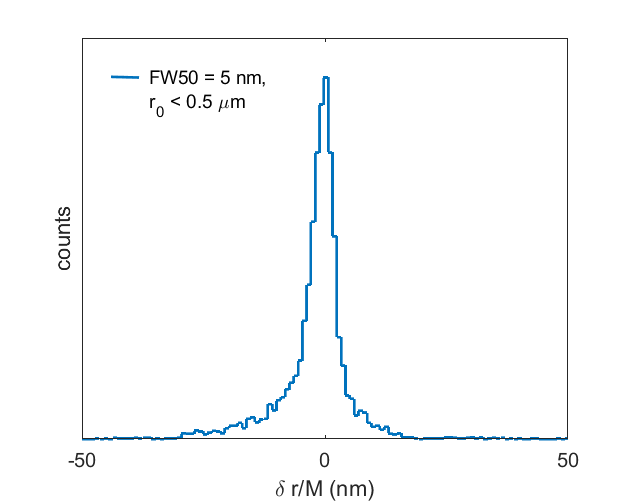}(b)
    \caption{(a) Image plane deviations as a function of initial angles for the Gaussian spatial illumination / uniform angular distribution beam case. (b) Histogram of the projected distribution within the 0.5 $\mu$m object space core.}\label{fig:gausbeam}
\end{figure*}



Trading off temporal resolution for spatial resolution may be undesirable for many time-resolved TEM application. In this section, we investigate the possible resolution improvements that can be attained by exchanging functional dependence of momentum space with real space. In practice, this can be done by using the condenser lens to refocus the beam after an overfilled aperture so that the illuminating momentum distribution is uniform, while the spatial distribution at the sample becomes Gaussian. If the spatial and momentum space configurations are exchanged in this way, then the second derivative of the space charge density modified to:
\begin{equation}
\rho^{(2)}(z)=-\frac{Q\exp{(-p^{-2}/2) }}{2 \pi\sigma_{r}^4C^4 L_{b}}
\end{equation}

In this case, $p=\frac{\sigma_rC(z)}{\theta_0S(z)}$, where $\sigma_r$ and $\theta_0$ are the gaussian rms spot size and hard-edge angular deflection of the object plane distribution. Consequently, the aberration coefficients are also modified:
\begin{equation}
C_e^{(p)}=-\frac{K}{8\sigma_{r}^3\theta_0}\int_0^L\frac{\exp{\left(-p^{-2}/2\right)}}{p}dz
\end{equation}\begin{equation}
C_e^{(q)}=-\frac{K}{8\sigma_{r}^2\theta_0^2}\int_0^L\frac{\exp{\left(-p^{-2}/2\right)}}{p^2}dz
\end{equation}
\begin{equation}
C_e^{(r)}=-\frac{K}{8\sigma_{r}\theta_0^3}\int_0^L\frac{\exp{\left(-p^{-2}/2\right)}}{p^3}dz
\end{equation}
\begin{equation}
C_e^{(s)}=-\frac{K}{8\theta_0^4}\int_0^L\frac{\exp{\left(-p^{-2}/2\right)}}{p^4}dz
\end{equation}

This particular object plane distribution gives rise to a stronger non linear dependence on $r_0$. In addition, particles that start far from the beam core and are initially converging towards the core will have the largest image plane deviation. Conversely, particles on the outskirts, diverging from the beam are deflected much less by collective forces.

This behavior is shown in Fig. \ref{fig:gausbeam}, where GPT simulation results are presented. The parameters for this simulations are from Table \ref{parameters}, where instead of a Gaussian beam divergence, a uniform angular distribution of up to 3~mrad is used at the object plane. The perveance and peak dose are the same as the beam simulated in Fig. \ref{Eq:spherical_aberration}. In Fig. \ref{fig:gausbeam}(a) output image plane deviations are shown color coded with respect to the initial position, $r_0$. Notably, the particles farther from the optic axis, moving toward the core, have the largest image plane deviation. The sparse residual smearing uncorrelated with $r_0$ is instead correlated with $r_0\theta_0'$. Plotted along with the output deviations is the predicted spherical aberration curve, which is found in good agreement with GPT outputs. In Fig. \ref{fig:gausbeam}(b), the histogram of the core outputs within a 0.5 $\mu$m offset from the optical axis is shown and results in FW50 spatial resolution of $5 nm$, nearly a factor of three better than the corresponding uniform/gaussian phase space distribution case, indicating that potentially large gains can be obtained by properly shaping the illumination in single shot time-resolved TEMs.

\begin{figure}[ht]
\includegraphics[scale=0.5]{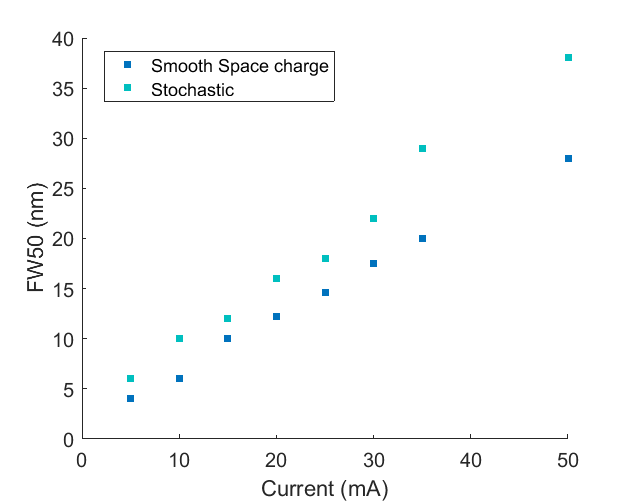}
    \caption{FW50 comparison between 'spacecharge3Dmesh' and 'spacecharge3D' models in GPT.}\label{fig:stoch}
\end{figure}
\section{Stochastic Scattering}
To provide better context for the analytical estimates discussed in the paper, we perform a final simulation campaign with GPT exploiting the full space charge 3D algorithm which takes into account the binary interaction between the particles to compare the image plane deviations obtained with our smooth space charge model. In order to keep computational times under control and limit the number of particles in the simulation, we only consider 100 fs slices with the same average current of the full beam. 
The reduced number of particles means that the Rose criterion would not be satisfied, but we can assess the spatial resolution considering the FW50 of the image plane deviations. Due to the relativistic dilation of distances in the beam rest frame, aspect ratios still remain very large, and the space charge field can be still be described on average.

The results are shown in Fig. \ref{fig:stoch}, where we compare the FW50 using two different space charge algorithms in GPT as a function of beam current. The smooth space charge algorithm solves the Poisson equation for a meshed charge distribution, and the stochastic one that takes into account all particle-to-particle interaction. The number of particles is adjusted in each simulation so that each macroparticle represents a single electron.  Parameters in Table \ref{parameters} were used to generate the beams used in simulation. The data in both cases is dominated by the linear contribution due to the smooth space charge effects. Only a small increase of the FW50 is observed as due to the binary interactions in the current range simulated. While for lower beam current, the binary interaction is proportionally more important and ultimately sets the limit for the spatial resolution, this plot confirms that compensating or correcting for the smooth space charge effects is a significant challenge in the development of single shot time-resolved TEMs. 

\section{Summary}

In summary, taking advantage of the Green's function technique to solve the driven Hill equation, we evaluate the smooth space charge-induced aberrations in a single shot time-resolved TEM and derived useful analytical expressions to estimate the spatial resolution as a function of beam parameters, energy and beam current. 
 
Our analysis is strictly valid in the perturbative regime, that is when the space charge non linear effects are small compared to the linear forces acting on the beam, but has been confirmed by benchmarking the results against full particle tracking simulations. The formulas allow quick estimates for the tradeoffs between spatial and temporal resolution in single shot time-resolved TEMs as shown in Section IV. It is shown that single nm spatial resolution could be achieved using relativistic beam energies, tens of ps long bunches and well-engineered spatial distributions.

In particular, in order to fully reap the benefits associated with the relativistic beam energy, it is important to develop electron optic technology which would allow to maintain very short focal lengths for the elements in the column. 

More complex optical setups involving quadrupole lenses or even multipole lenses could be employed to further reduce the smooth space charge aberration effects. The resolution limit of the instrument will then be set by the stochastic blurring associated with information-lossy coulomb collisions between the electrons of the beam in the column after the sample. 

We expect this work to elucidate the potential in single shot time resolved TEMs and guide the developments of future instruments for time-resolved microscopy applications. 

\section{Acknowledgments}
A lot of this work is based on D. Cesar's Ph.D. dissertation. We sincerely thank him for the enlightened discussions on the subject and for leading the way in setting up the problem. This work was supported by DOE STTR grant No. DE-SC0013115 and by by the National Science Foundation under STROBE Science and Technology Center Grant No. DMR-1548924. 

\section{Appendix}

\subsection{Gaussian Space Charge Aberration}
In the paper we discuss the cases of uniform/gaussian and gaussian/uniform illumination. Another important case is when the phase space charge density is a purely gaussian (both in momentum and position space). For an initially uncoupled beam we can write:
\begin{equation}
\rho=\frac{\lambda\exp\left(\frac{-r^2}{2(\sigma_{\theta}^2S^2+\sigma_r^2C^2)}\right)}{2\pi(\sigma_{\theta}^2S^2+\sigma_r^2C^2)}
\end{equation}
The on axis second derivative is found to be:
\begin{equation}
    \rho^{(2)}(z)=-\frac{\lambda}{2\pi\sigma_{\theta}^4S^4(1+p^2)^2}
\end{equation}
where in this case $p=\frac{\sigma_rC(z)}{\sigma_{\theta}S(z)}$. The aberration coefficients are then given by:
\begin{equation}
C_e^{(p)}=-\frac{K}{8\sigma_{r}^3\sigma_{\theta}}\int_0^L\frac{p^3}{(1+p^2)^2}dz
\end{equation}\begin{equation}
C_e^{(q)}=-\frac{K}{8\sigma_{r}^2\sigma_{\theta}^2}\int_0^L\frac{p^2}{(1+p^2)^2}dz
\end{equation}
\begin{equation}
C_e^{(r)}=-\frac{K}{8\sigma_{r}\sigma_{\theta}^3}\int_0^L\frac{p}{(1+p^2)^2}dz
\end{equation}
\begin{equation}
C_e^{(s)}=-\frac{K}{8\sigma_{\theta}^4}\int_0^L\frac{1}{(1+p^2)^2}dz
\end{equation}
The last term in these expression is responsible for the space charge spherical aberration. The integrand can only be smaller than 1. For optics and initial beam parameter are such that $p$ is small along most of the column (i.e. the spot size is dominated by the initial beam divergence), the integral can be safely approximated to $L$ so that the space charge spherical aberration of this case is similar to the uniform illumination/ gaussian angular distribution case.  
\bibliographystyle{apsrev4-1}
\bibliography{scaberr}

\end{document}